# Overcoming optical contrast limit of mesoscale light focusing by means of retrograde-reflection photonic nanojet

YURY E. GEINTS,[1,*] ALEXANDER A. ZEMLYANOV,[1] IGOR V. MININ[2,3], OLEG V. MININ[2,3]

[1]V.E. Zuev Institute of Atmospheric Optics, 1 Acad. Zuev square, Tomsk, 634021, Russia
[2]Tomsk Polytechnic University, Lenina 36, Tomsk, 634050, Russia
[3]Tomsk State University, Lenina 30, Tomsk, 634050, Russia
[*]Corresponding author: ygeints@iao.ru

The physical origin of subwavelength photonic nanojet in retrograde-reflection mode (retro-PNJ) is theoretically considered. This specific type of photonic nanojet emerges upon sequential double focusing of a plane optical wave by a transparent dielectric microparticle located near a flat reflecting mirror. For the first time to the best of our knowledge, we report a unique property of retro-PNJ for increasing its focal length and intensity using microparticles with the optical contrast exceeding limiting value (n > 2) that fundamentally distinguishes retro-PNJ from regular PNJ behavior. This may drastically increase the trapping potential of PNJ-based optical tweezers.

Photonic nanojet (PNJ) is a specific near-field focusing region of the optical wave illuminating a dielectric microparticle with the dimensions of several radiation wavelengths (mesoscale particle) [1]. As mesoscale light structures, PNJs allow the electromagnetic field to be concentrated at sub-diffraction spatial scales [1, 2] that finds wide application in various fields of science and technology, including laser modification of surfaces [3], nano- and microscopy [4], fluorescence enhancement [5], nanolithography [6] as well as optical trapping (OT) of various nanoobjects and even atoms [7, 8]. When mesoscale dielectric particles are used in the OTs as a focusing device, they are usually placed in a liquid medium were the refractive index contrast of the particle and medium is less than two [1,2,7,8]. Importantly, among other types of OTs, PNJ-based traps can effectively capture both dielectric and metallic particles, considering the optical radiation pressure is much higher for metallic particles that is challenging for conventional Gaussian tweezers. Thus, PNJ is a promising candidate for manipulating and trapping nanoparticles at mesoscales.

In conventional PNJ-based OTs, the concentrated photonic flux is formed by an optical wave in the "transmission mode", when the area of field spatial localization is originated in the shadow part of the focusing microparticle [2]. Recently, for the first time the PNJ generation in the "reflection mode" was reported in [9,10] during electromagnetic wave reflection from a metal flat mirror attached to a transparent dielectric cubic [9,10] or hemispherical microparticle [11]. In this case, the so-called *retro*-PNJ is modulated by a standing wave resulting from the constructive interference of the incident and reflected waves [9-12]. Similar light structures in the form of standing waves arise in the interaction of two counter-propagated PNJs [13], as well as when the microparticle is placed on a dielectric screen with relatively high dielectric permeability [14,15].

Worthwhile noting, all the optical trapping schemes considered above demand the specific condition to be always satisfied saying that the optical contrast $n$ of the microparticle with respect to the environment must be less than a certain threshold value $n^*$, which depends on particle size but does not ever exceed two [1, 2, 7-16]. Only under this condition, the region of mesoscale field localization is located outside the particle. Indeed, from the geometrical optics consideration one knows the formula for the focal length of a ball-lens of radius $R$: $f = nR/2(n-1)$. It is clear that for $n > n^* = 2$ the focus of such a lens should be inside the microsphere ($f < R$). On the mesoscale, the situation with $n > n^*$ means that for high-contrast microparticles the emerging photonic jet is located mostly inside the particle that significantly reduces its external intensity maximum. This circumstance limits the range practical application of PNJ-based traps to the cases of low-contrast particles ($n < 1.2$) which produce only moderate field intensity enhancement.

In this work, for the first time to the best of our knowledge we show the way to overcome this contrast limitation by means of retrograde-reflection PNJ (*retro*-PNJ). As mentioned above, this specific *retro*-PNJ is formed upon plane wave reflection from a flat metal mirror placed near a nonabsorbing microparticle [9-12], e.g., a glass microcylinder. We discuss the physical mechanism of *retro*-PNJ formation by high-contrast particles and show that this type of PNJs is situated outside the microparticle even above the "critical" refraction index contrast $n > n^*$. Meanwhile, in such conditions the *retro*-PNJ increases its longitudinal extension up to several optical wavelength while maintaining sub-wavelength transverse dimensions. Such jet parameters cannot be obtained under conventional scheme of PNJ generation in high-contrast particles.

The mesoscale photonic structure under study, as an example, is a dielectric microcylinder with variable refractive index (RI) $n_1$, immersed in water with refractive index $n_0$= 1.33. All dielectric materials are assumed to be nonabsorbing for optical radiation. In the usual PNJ-generation scheme (Fig. 1a), a concentrated optical flux occurs near the shadow region of the microcylinder exposed to an optical radiation. In the scheme with retro-reflection (Fig. 1b), a thin high-reflective plate is placed near one side of the cylinder and serves as a retrograde reflection mirror for the light wave illuminating the microcylinder from the opposite side. This can be a metal mirror, such as metal foil [12], or a high-contrast (Si) dielectric plate [15]. For definiteness, in the simulations the diameter of the microcylinder is fixed at 4λ.

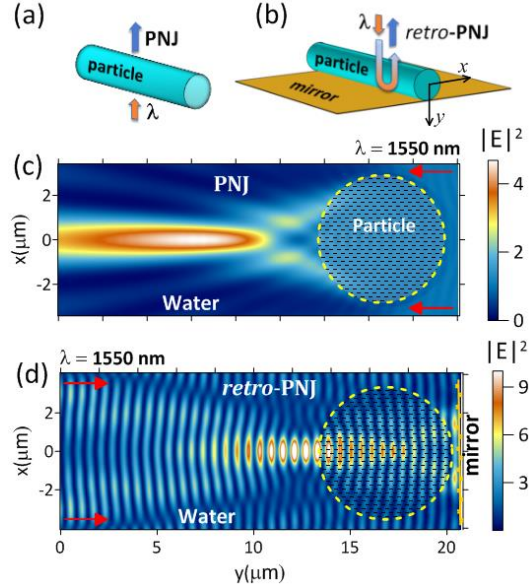

Fig. 1. (a,b) Schematics of (a) PNJ and (b) *retro*-PNJ formation by a dielectric microcylinder exposed to an optical wave λ; (c,d) optical intensity $|E|^2$ distribution of (c) PNJ and (d) *retro*-PNJ generated by a glass cylinder ($n_1$=1.45) immersed in water.

The simulations of the near-field distribution are carried out on the basis of the numerical solution of the Helmholtz equation for vectorial electromagnetic field using the finite elements method (FEM) implemented in COMSOL Multiphysics commercial solver (version 5.1). The two-dimensional geometry of the problem in *xy*-coordinate plane is used, and along the *z*-axis the photonic nano-structure is considered infinite. The microcylinder is assumed to be illuminated by a plane *p*-polarized optical wave with telecommunication wavelength λ = 1550 nm. The maximal size of the spatial triangular-mesh is chosen as 5 nm inside 70 nm-thick gold mirror-plate and λ/15 in the remaining calculation domains. The details of near-field scattering numerical simulations via COMSOL Multiphysics software can be found elsewhere [17, 18].

Figures 1c, 1d exemplify the distribution of optical wave intensity $|E|^2$ at the diffraction at a glass microcylinder located in water ($n$ = 1.09 [15]) in the cases of conventional "transmission" PNJ and *retro*-PNJ. As seen, in both cases in the near-field area of a particle a focal region with enhanced intensity is formed. The presence of a flat metal mirror near the shadow side of the microcylinder in the retro-reflection mode (Fig. 1d) causes the modulation of PNJ bell-shaped longitudinal profile by a standing wave with the antinode period of $\lambda/2n_0$ [12,13]. Importantly, the *retro*-PNJ exhibits an almost twofold intensity increase in comparison with the conventional PNJ.

It is well known that in the classical scheme of PNJ generation by a dielectric particle, the increase of particle contrast inevitably leads to photonic jet shortening and its moving toward the shadow particle edge [1,2,7-16]. In Figs. 2a and 2b we show the calculated dependence of the peak intensity $E_{max}^2 = \max_{xy} |E|^2$ and focal distance *f* of PNJs in transmission and retro-reflection modes by varying the relative refractive index *n* (optical contrast) of microcylinder. In the simulations, PNJ focal distance *f* is measured from the nearest to the photonic jet particle surface to the position of external field intensity maximum.

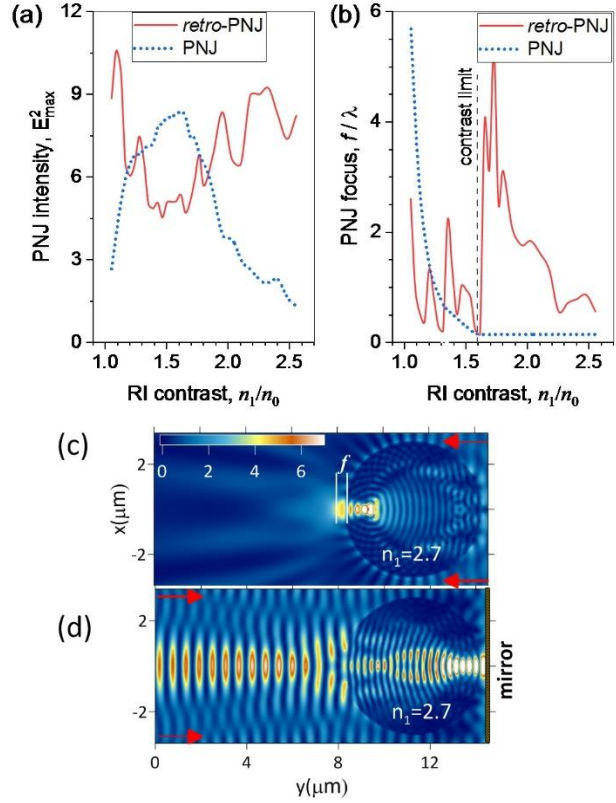

Fig. 2. (a) Maximal intensity $E_{max}^2$ and (b) focal distance $f$ in conventional PNJ and *retro*-PNJ for different optical contrast $n = n_1/n_0$ of microcylinder; (c, d) intensity distribution in PNJ (c) and *retro*-PNJ (d) at supercritical contrast $n > 2$.

As seen, for the conventional PNJ the increase of particle contrast $n$ first leads to an increase in intensity to the value of $E_{max}^2 = 8.2$, and then, at $n > n^* \approx 1.61$, the intensity gradually decreases almost to the initial unperturbed value. In this case, the distance $f$ of PNJ intensity maximum monotonically decreases with particle RI-contract increase. As a result, in the range of supercritical $n$-values, the focusing area is localized almost completely inside the particle volume. PNJ "docks" to the microcylinder and is looks like a short exponentially decaying bulb in the geometric shadow of the particle (Fig. 2c).

However, the PNJ in retrograde reflection mode demonstrates a fundamentally different parameters trend. Here, the increase of microcylinder contrast first decreases the PNJ intensity, but after overcoming the limiting value, in our case $n^* \approx 1.61$ (fig.2b), the PNJ intensity on the contrary increases. This growth is non-monotonous and possesses a resonance character which resembles the standing wave formation in the Fabry-Perot resonant cavity.

Noticeable, for particle contrasts close to the critical value $n^*$, *retro*-PNJ focal length changes sharply, from $f \sim \lambda/4$ for $n = 1.6$ to $f \sim 5\lambda$ at $n = 1.8$. At the same time, conventional PNJ in this contrast range has very short focal length of about $\lambda/5$ and its spatial extent does not exceed a half-wavelength [1, 2]. Further increase of particle contrast brings again the center of the *retro*-PNJ closer to the particle edge but $f$ still remains greater than $\lambda$ even for the contrast $n > 2$ (Fig. 2d). Meanwhile, in the supercritical region, $n > n^*$, the retro-jet intensity undulatory increases practically to the value observed for low-contrast particles [15], while maintaining subwavelength transverse size that is impossible for conventional PNJs.

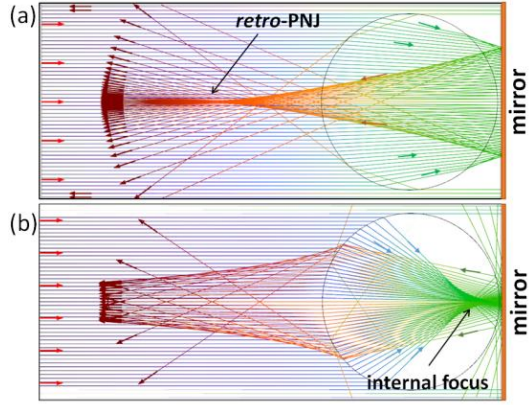

Fig. 3. GO ray tracing of retro-PNJ region formed by a microcylinder with $n = 1.09$ (a) and $n = 2$ (b). Different ray path-length is rainbow colored for visibility.

For understanding the physical causes of this abnormal effect in the optics of photonic jets, we suggest a geometrical optics (GO) treatment of this problem. Strictly speaking, the geometrical optics on mesoscales ($R \propto \lambda$) is not applicable. However, such a simplified approach provides for qualitative physical picture and clearly indicates the physical interpretation of the phenomenon in question.

Within the framework of GO approximation, the photonic jet is a specific region bounded by the external ray caustics which are formed by focusing light rays at a particle. Thus, a PNJ is a result of light rays concentration due to multiple refractions at optically contrasting boundaries of the physical media.

The results of geometric ray tracing procedure are presented in Figs. 3a, b for two values of particle contrast. As seen, *retro*-PNJ for particle with $n = 1.09$ (Fig. 3a) is formed by a ray family comprised from specific incident ray bundle which is focused by the particle into a wide area (due to low contrast) on the mirror and then reflected backward and refocused by particle. As a result, the length of the focal caustic in this case is short and PNJ is originates close to the shadow surface of the microcylinder.

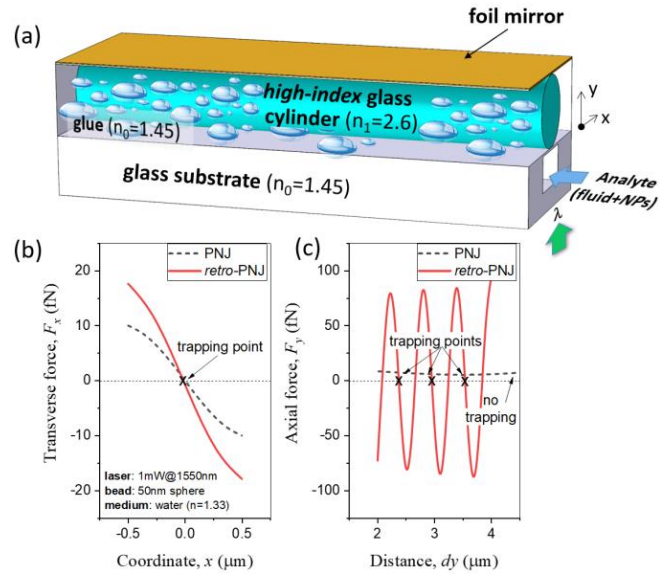

Fig. 4. (a) Schematic of proposed multi-position *retro*-PNJ tweezer; (b, c) net optical forces acting on 50 nm gold bead along *x*-axis (b) and *y*-axis (c) for two types of PNJs.

The initial stage of retro-PNJ formation for supercritical contrast $n = 2$ (Fig. 3b) is similar to the previous case except the fact that the primary ray focus is not behind the mirror but rather lays inside the particle volume near its shadow edge. As a result, a narrow high-divergence beam of rays emerges from the inner focus. After reflection at the mirror, this divergent beam hits again the rear (shadow) surface of the particle, and then after two consequent refractions becomes convergent. Since such focusing is quite weak, the external focal area, i.e. *retro*-PNJ, is located farther from the microcylinder than for the particle with low optical contrast in Fig. 3a. However, because of this loose focusing the intensity in the external focus is lower than in the case of very low contrast (Fig. 2a). The intensity in *retro*-PNJ can be increased by sharpening the ray focusing via increasing the microparticle contrast but this may shorten the focal distance.

Based on the analysis of intensity distribution in retro-PNJ, in Fig. 4a is presented one of the possible engineering design of a multi-position optical tweezer with distributed trapping region as applied to the integrated optofluidic devices.

The OT proposed is designed as an on-chip microfluidic device consisting of a glass base with a hollow rectangular channel fabricated near one of the matrix edges. The channel shape can be arbitrary, for example, rectangular as in the figure, since it has weak effect on the distribution of optical trapping forces in the channel. On top of the glass matrix, above the channel, a focusing microcylinder is mounted. It can be fabricated from some type of infrared glass (BD-2, IRG-22 [21]) with high refractive index, providing supercritical contrast value $n > 1.6$. Above the cylinder a metal foil is mounted that acts as a retro-reflecting mirror. The entire structure can be fixed with a transparent glue having refractive index close to that of glass substrate and is illuminated upwards by a laser with, e.g., telecommunication wavelength 1550 nm.

The liquid analyte containing the nanoparticles (NPs) to be manipulated is flushed through a microfluidic channel. As soon as NPs enter the trapping area of the retro-PNJ they occupy multiple potential wells of the *retro*-PNJ trap under the action of optical capture forces. While moving in a liquid flow through the channel, the randomly distributed NPs are ordered and form the lined nanoparticles "beamlets", which can then be further processed and analyzed on demand [22].

Net optical forces [7, 13] acting on a 50-nm Au-sphere placed in proposed retrograde PNJ-based trap are presented in Figs. 4b,c. In the calculations, a laser beam with the power 1 mW is focused in the spot with the dimensions of $10 \times 10$ µm$^2$. Test nanoparticle is immersed in water. For comparison, we also calculated the trapping forces for conventional PNJ-trap, which is engineered as in Fig. 4a but without a mirror and with laser exposure directed downwards. Additionally, we replaced the focusing microcylinder by a lower-contrast glass fiber with $n_1 = 1.74$ to obtain similar PNJ focal distance $f = 3\lambda$ as in *retro*-PNJ (see, Fig. 2b). For each OT a gold bead is placed in the region of field intensity maximum, where its position can vary within certain limits.

As can be seen from Fig. 4b in the comparison with conventional PNJ, retro-PNJ shows noticeably larger transverse restoring optical force $F_x$ with the amplitude ∼ 20 fN within $\lambda/2$ displacement. This indicates both large gradients of field amplitude and a multiple increased intensity on the trap axis.

Axial optical forces $F_x$ versus bead displacement $dy$ from microcylinder edge are shown in Fig. 4c. Here, the most indicative is that 50 nm metal particle was too large for trapping by conventional PNJ. In this case, the optical trapping is not realized due to a strong imbalance in favor of the scattering force that tends to pull out the nanoparticle from the trapping point in the direction of light beam propagation. However, the proposed trap on the standing wave, due to the obvious dominance of the gradient component of net optical force, shows a reliable capturing of a 50 nm gold sphere in several positions (marked by crosses in the figure 4). Obviously, these positions are close to antinodes of longitudinal intensity distribution. Moreover, the retro-PNJ trap has stronger trapping ability in this direction.

In summary, we considered the specific regime of near-field focusing of an optical wave by a dielectric microcylinder on top of a metal reflective mirror – the *retro*-PNJ. We found that in a number of cases *retro*-PNJ has higher intensity than conventional PNJ. We reveal a unique feature of *retro*-PNJ for overcoming the contrast limit ($n < 2$) known in mesoscale optics for plane wave focusing by a dielectric microparticle [1,2,16]. Thus, by using, e.g. a microcylinder with supercritical optical contrast, $n > 1.61$ for generating *retro*-PNJ, one can increase the jet intensity and its length rather than their reduction as with conventional PNJ.

We propose relatively simple technical implementation of the optical tweezer based on *retro*-PNJ effect which can find practical application as a chip-on-flex microfluidic device for multi-position optical sorting and beaming of target atoms or nanoobjects [23, 24].

## Funding

**Disclosures**. The authors declare no conflicts of interest.